\documentclass[12pt]{article}
\usepackage[utf8]{inputenc}
\usepackage{comment}
\usepackage{enumitem}
\usepackage{setspace}
\usepackage{siunitx}
\usepackage[margin=1in]{geometry}
\usepackage[dvipsnames]{xcolor}
\usepackage[english]{babel}
\usepackage{caption}
\usepackage{subcaption}
\usepackage[sort]{natbib}

\usepackage{lineno,blindtext,algorithmic,graphicx,amssymb,amsmath}
\usepackage[colorinlistoftodos]{todonotes}
\usepackage{threeparttable}
\usepackage{authblk}

\begin{document}

	\title{Comparison of Effect Size Measures for Mediation Analysis of Survival Outcomes with Application to the Framingham Heart Study}
	\author[1,2]{Bin Shi}
    \author[1*]{Xuelin Huang}
	\author[1*]{Peng Wei}
	\affil[1]{Department of Biostatistics, The University of Texas MD Anderson Cancer Center, Houston, Texas, USA}
	\affil[2]{Department of Biostatistics and Data Science,The  University of Texas Health Science Center at Houston,Houston, Texas, USA}
	\affil[*]{Correspondence should be addressed to: Peng Wei, PhD, pwei2@mdanderson.org, or, Xuelin Huang, PhD, xlhuang@mdanderson.org}
	
\date{}
\maketitle
\vspace{-1.2cm}
\doublespacing

\begin{abstract}
There is an increasing trend of research in mediation analysis for survival outcomes. Such analyses help researchers to better understand how exposure affects disease outcomes through mediators. However, due to censored observations in survival outcomes, it is not straightforward to extend mediation analysis from linear models to survival outcomes.
In this article, we extend a mediation effect size measure based on $R^2$ in linear regression to survival outcomes. Due to multiple definitions of $R^2$ for survival models, we compare and evaluate five $R^2$ measures for mediation analysis. Based on extensive simulations, we recommend two $R^2$ measures with good operating characteristics. We illustrate the utility of the $R^2$-based mediation measures by analyzing the mediation effects of multiple lifestyle risk factors on the relationship between environmental exposures and time to coronary heart disease and all-cause mortality in the Framingham Heart Study.
\end{abstract}
\textbf{Keywords}: Mediation analysis; R-squared measure; Cox survival analysis; Indirect Effect.

\section{Introduction}
There is increasing popularity of mediation analysis in medical research, especially mediation with time-to-event outcomes \citep{Lapointe2018}. In our motivating example, we focus on cardiovascular disease, including coronary heart disease (CHD), from the Framingham Heart Study (FHS). It is widely accepted that evaluating well-established lifestyle risk factors on CHD events plays an important role in controlling and preventing CHD occurrence. Such risk factors typically include  lipid profile (i.e., total cholesterol (TC), low density lipoprotein cholesterol (LDL-C), high density lipoprotein cholesterol (HDL-C), and triglycerides (TG)), blood pressure (BP), obesity, diabetes, and cigarette smoking \citep{Wilson1998}. There is growing evidence showing that these risk factors lie in the causal pathway between environmental exposure and downstream CHD events or mortality outcomes \citep{Buttar2005, Sharif2019}. Thus, it is of great interest to understand and estimate the additional proportion of variation in disease outcomes explained by environmental risk factors through these putative mediating pathways. Mediation analysis is able to meet such needs by helping researchers to explain why or how environmental exposures affects disease outcomes \citep{MacKinnon2008,Sampson2018}. 

Despite increasing research trends in the area of mediation analysis, most studies focus on a single mediator, linear regression, or causal interpretation in simple mediation structures due to the lack of preferred, flexible effect-size measures \citep{Vandenberghe2018}. Here, we would like to extend mediation analysis from linear outcomes to survival outcomes using an $R^2$ measurement, instead of the commonly used product-based mediation effect size measures \citep{VanderWeele2011,Huang2017}. When considering other existing effect size measures, we prefer the use of $R^2$ for survival outcomes for the following reasons: 1) easier to extend to high-dimensional space \citep{Fairchild2009, Yang2019}; 2) high computing efficiency and accuracy when p is larger than 2, of high dimensions, or the mediators are in highly correlated structures \citep{Yang2019}; 3) the component-wise mediation effect-size measures may be positive or negative, leadings to cancellation in the estimation of the total mediation effect via the marginal mediation effect measures, such as the sum of the product measures. The $R^2$ measurement avoids such issues \citep{Fairchild2009, Yang2019}.

The chosen $R^2$ measure for survival outcomes also poses another layer of challenges to mediation analyses. There are multiple proposals for $R^2$ measures in survival models \citep{HonerkampSmith2016, ChoodariOskooei2012a, ChoodariOskooei2012b}, and all of the proposed $R^2$ measures can be classified into three groups: explained variation, explained randomness, and predictive accuracy. However, it is not clear which $R^2$ measure will serve best as a tool for mediation analysis for Cox regression models.

According to the properties of a good measure of $R^2$ suggested by \cite{Royston2006}, we chose five different $R^2$ measures to evaluate mediation effect performance through simulation studies and real data settings. We recommend two of the five that are better effect size measures for multiple-mediator models under the Cox proportional hazards (PH) framework from simulation studies. In the FHS data analysis, we found that six lifestyle risk factors as mediators altogether can explain more than 80\% of the exposure variables-associated CHD risk. Under the same setting, however, they can only explain 5\% - 14\% of all-cause mortality risk associated with the same set of exposure variables.

The rest of the paper is organized as follows. Section 2 presents details of the statistical models that we addressed, specifically the mediation analysis under the Cox PH regression models. Section 3 presents simulation studies. Section 4 provides the results of the real data analysis. Section 5 concludes with a discussion.

\noindent
\section{Methods}

\subsection{Extension of Mediation Analysis for Continuous Outcomes to Survival Outcomes}
A large body of literature describe mediation analyses under the linear regression setting for continuous outcomes. In this research, we focus on the common setting with survival outcomes in clinical practice: Cox PH regression models.
\subsubsection{Single-Mediator Analysis}
A single-mediator model is the simplest case of mediation analysis, where the association between an independent variable $X$ with a dependent variable, time-to-event $T$, is accounted ``partially" or ``entirely"  through the third variable, a mediator $M$ (Figure~\ref{f1}). The single-mediator model under a Cox regression setting can be described using the following equations:

\begin{figure}[h!]
\begin{center}
\includegraphics[width=10cm,height=6cm]{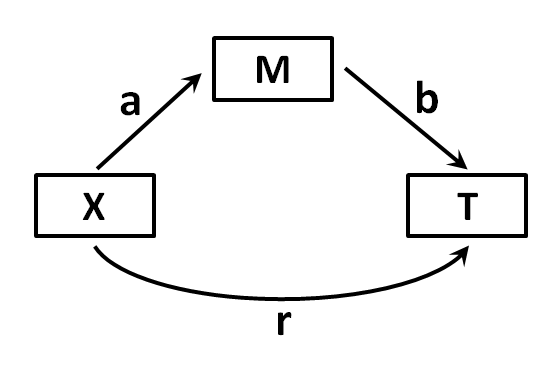}
\end{center}
\caption{Single-Mediator Models}
\label{f1}
\end{figure}

\vspace{-0.5cm}
\begin{align}
& \lambda(t|X) = \lambda_0^{(X)}(t) \exp{(cX)} \label{eq: X_single},\\
& \lambda(t|X,M) = \lambda_0^{(XM)}(t) \exp{(rX + bM)} \label{eq: XM_single},\\
& \lambda(t|M) = \lambda_0^{(M)}(t) \exp(dM) \label{eq: M_single},\\
& M = e + aX + \epsilon \label{eq: med_single},
\end{align}
\noindent
where $e$ is an intercept, and $a,b,c,d$ and $r$ are regression coefficients. Note that $\lambda_0^{(X)}(t)$, $\lambda_0^{(XM)}(t)$ and $\lambda_0^{(M)}(t)$ are baseline hazard functions for the three Cox PH models, respectively. $\exp{(r)}$ is the direct effect, $\exp{(a*b)}$ is the indirect effect, and $\exp{(c)}$ is the total effect. There are two equivalent ways to compute the mediation (indirect) effect: product and difference measures \citep{VanderWeele2011}. The product measure is to obtain the product of the coefficients $a$ and $b$, and the difference measure is to subtract the effect $r$ from total effect $c$, where $c = ab + r$, originally proposed in linear regression models by \cite{Baron1986}. Both measures are unstandardized effect size measurements of mediation. Normalized by the total effect, the proportion of the indirect effect (proportion mediated) can be calculated by the following equations:
\begin{align*}
& \text{Product-based proportion mediated}  = \frac{\exp(a*b)}{\exp(c)}, \\
& \text{Difference-based proportion mediated} =  \frac{\exp{(c - r)}}{\exp{(c)}} = \exp{(- r)}.
\end{align*}
In addition to reporting the proportion mediated effect sizes, there have been many other mediation effect size measures proposed in the literature. Partially standardized indirect effect and fully standardized indirect effect have been evaluated for both single-mediator models and two-mediator models, which both have satisfactory relative bias levels and high efficiency \citep{Miocevic2017}. The proportion mediated and ratio mediated effect sizes are most commonly seen in the literature; however, both of them require large sample sizes to obtain stable estimates \citep{MacKinnon2008}. All of these measures face problems in inconsistent mediation models and multiple-mediator models where $a*b$ and $c$ have opposite signs, which is known as suppression via effect-size cancellation by the different signs of coefficients $a$ and $b$. One option to deal with the different signs of $a, b$ and $c$ is to take the absolute value of each before computing the proportion mediated \citep{Alwin1975}. However, this is hard to interpret.

$R^2$ measures, the explained portion of variance in an outcome variable by the indirect effect, were originally proposed from commonality analysis by \cite{Fairchild2009}, and they have been developed only for continuous outcomes \citep{Yang2019}. Under Cox regression models, we propose to compute the $R^2$ mediation effect along the same line as in linear regression:

\begin{equation}
R^2_{med} = R^2_{T,M} + R^2_{T,X} - R^2_{T,MX},
\end{equation}
\noindent
where $R^2_{T,M}$ is the proportion of variance in $T$ that is explained by $M$, $R^2_{T,X}$ is the proportion of variance in $T$ that is explained by $X$, and $R^2_{T,MX}$ is the proportion of variance in $T$ that is explained by $M$ and $X$. $R^2_{med}$ can be interpreted as the amount of variance in $T$ that is explained by $M$, specific to the mediated effect if no other unmeasured confounders exist. The counterpart of standardized $R^2$ measures is called the shared over simple effect (SOS) \citep{Lindenberger1998}, and calculated via the formula: $SOS = R^2_{med}/R^2_{T,X}$.

\subsubsection{Multiple-Mediator Analysis}
\begin{figure}[h!]
\begin{center}
\includegraphics[width=10cm,height=8cm]{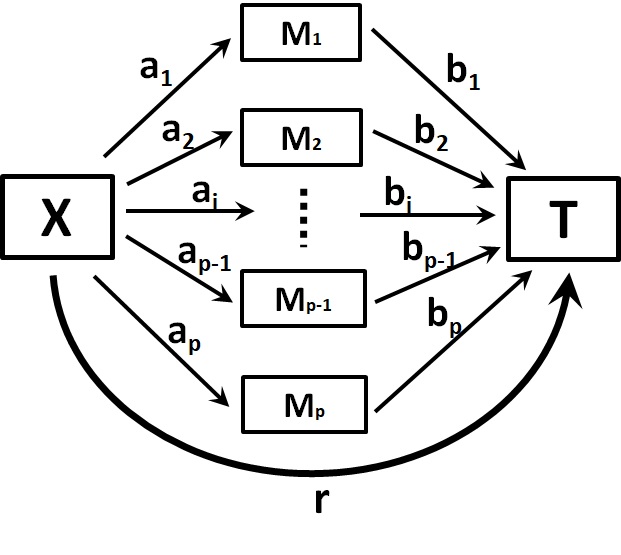}
\end{center}
\vspace{-0.5cm}
\caption{Multiple-Mediator Models}
\label{f02}
\end{figure}

Figure~\ref{f02} illustrates a model for multiple-mediator analysis, which can be written as follows,

\begin{align}
& \lambda(t|X) = \lambda_0^{(X)}(t) \exp{(cX)} \label{eq: X_multiple},\\
& \lambda(t|X,M) = \lambda_0^{(XM)}(t) \exp{(rX + \sum b_j M_j)} \label{eq: XM_multiple},\\
& \lambda(t|M) = \lambda_0^{(M)}(t) \exp{(\sum d_j M_j)} \label{eq: M_multiple},\\
& M_j = e_j + a_jX + \epsilon_j \label{eq: med_multiple},
\end{align}
where $a_j$, $b_j$, $r$, $d_j$, and $c$ are regression coefficients, $e_j$ is an intercept for $j = 1, 2, \cdots, p$, and $p$ is the number of mediators. Note that $M_j$ is regressed on $X$ for each mediator. $\lambda_0^{(X)}(t)$, $\lambda_0^{(XM)}(t)$ and $\lambda_0^{(M)}(t)$ are baseline hazard functions for the three Cox PH models, respectively. Similarly to the single-mediator models, the mediation indirect effect can be calculated through the exponential summation of the products: $\exp(\sum (a_jb_j))$ or exponential difference: $\exp(c - r)$. However, they become more complicated by the addition of multiple mediators, because the directions of $a_{j}*b_{j}$ and $r$ can be very likely different. Standardized proportion mediated measures have been defined in multiple-mediator models as follows:
\begin{align*}
& \text{Product-based proportion mediated}  = \frac{\exp(\sum a_j b_j)}{\exp(c)}, \\
& \text{Difference-based proportion mediated} =  \frac{\exp{(c - r)}}{\exp{(c)}} = \exp{(- r)}.
\end{align*}
Evaluation of the explained variance $R^2$ measure under multiple-mediator models is similar to that used in single-mediator models, and it is straightforward to extend $R^2$ to multiple-mediator models as follows:
\begin{align}
& R^2_{med} = R^2_{T,M} + R^2_{T,X} - R^2_{T,MX} \label{eq: R2medPH}, \\
& SOS = R^2_{med}/R^2_{T,X} \label{eq: SOSPH},
\end{align}
\noindent
where $R^2_{T,X}$, $R^2_{T,MX}$ and $R^2_{T,M}$ are based on models (\ref{eq: X_multiple}),  (\ref{eq: XM_multiple}), and (\ref{eq: M_multiple}), respectively. 
$R^2_{med}$ is interpreted as the difference of the variance of the dependent variable $T$ that is explained by the exposure $X$ ($R^2_{T,X}$: total effect), and the variance that can be explained by the exposure after removing the mediator mediated effect ($R^2_{T,MX}-R^2_{T,M}$: the so-called direct effect). The $R^2_{med}$ measure has many characteristics of a good measure of effect size. It is appealing to measure the total mediation effect of multiple-/high-dimensional mediators \citep{Yang2019}.  It has been shown to have stable performance for sample sizes $> 100$ \citep{Fairchild2009}. Its scale is appropriate for most of the given questions of interest,  and it increases as the mediation effect approaches the total effect. 
A major concern is that it can be negative, hindering its interpretation, for which we will give more detailed discussion in the Results and Discussion sections. Here we would like to evaluate performance of the $R^2_{med}$ measure under a multiple-mediator Cox regression setting ($n >> p$ and $p \le 10$). 

\subsection{Comparison of Different $R^2$ Measures for Multiple-mediator Mediation Analysis under the Cox PH Models} \label{R2prop}
Motivated by the definition of $R^2_{med}$ in the linear regression framework, we calculate mediated effects by $R^2_{med}$ measures for Cox PH models as defined in formula (\ref{eq: R2medPH}). In general, $R^2$ measures in survival models are calculated by the formula: $ R^2 = 1-\frac{Var(g(T)|Z)}{Var(g(T))}$, where $g$ is a link function.

\cite{Royston2006} suggested that a good measure of $R^2$ should have the following five properties:
(a) closely related to explained variation, which is ``equivalent" to linear regression analysis;
(b) approximate independence of the amount of censoring;
(c) $R^2$ increasing with the strength of association;
(d) nesting property: for two models, $M_1$ nested in $M_2$ should have $R^2 (M_1) < R^2 (M_2)$;
(e) availability of confidence intervals.

Based on the above properties, we chose five different $R^2$ measures to assess the performance of the corresponding $R^2_{med}$ mediation effect measure under the Cox PH modeling framework in simulation studies and real data analysis. Two of the five different $R^2$ measures are defined as follows \citep{Royston2006}:
\begin{align}
& R_n^2 = 1-\exp{\{- \frac{2}{n}(l_{\hat{\beta}} - l_0)\}} = 1- \exp{(-\frac{\chi^2}{n})} \label{R2n},\\
& R_k^2 = 1-\exp{\{- \frac{2}{e}(l_{\hat{\beta}}} - l_0)\} = 1- \exp{(-\frac{\chi^2}{e})} \label{R2k},
\end{align}
where $n$ is the sample size, and $e$ is the number of events. $l_{\hat{\beta}}$ denotes the maximized log likelihood under the working model with $\beta$, and $l_0$ is the maximized log likelihood of the null model. $ \chi^2 = 2(l_{\hat{\beta}} - l_0)$ is the likelihood ratio statistic for comparing between the model with $\beta$ and the null model without $\beta$. Equation (\ref{R2n}) is a classic measure of the proportion of explained variance of outcome dependent on the predictors in a general regression model. It was originally developed by \cite{Nagelkerke1991}, and we call it $R_n^2$ . Equation (\ref{R2k}) generalizes the $R_n^2$ measure for application to the Cox PH regression models, as done by O'Quigley et.al (2005). This measure was considered for computing explained variance in the context of censored survival data, instead of only considering sample size \citep{O'Quigley2005}. We call the measure $R_k^2$. The other three $R^2$ measures are defined as follows \citep{Royston2006, Heller2012}:
\begin{align}
& R_r^2 = \frac{A}{\frac{\pi^2}{6} + A} = \frac{R_k^2}{R_k^2 + \frac{\pi^2}{6}(1 - R_k^2) } \label{R2r},\\
& R_b^2 = \frac{B}{0.5772 + B} \label{R2b},\\
& R_w^2 = \frac{var(\textbf{Z}\boldsymbol{\beta})}{1 + var(\textbf{Z}\boldsymbol{\beta})} \label{R2w},
\end{align}
where $A = R_k^2/(1 - R_k^2)$, $\textbf{Z} = (X, \textbf{M})$ \citep{Royston2006}, $B = log(\frac{\sum_{i=1}^{n} exp(\boldsymbol{\beta}^T \textbf{z}_i)}{n})$, $\textbf{z}_i$ is an element of $\textbf{Z}$, $ i = 1, \cdots, n$, and $ \frac{\pi^2}{6} -1 = 0.5772$ \citep{Heller2012}.

Equation (\ref{R2r}) is a function of the $R_k^2$ measure, which was developed by Royston (2006) to achieve better performance in Cox PH regression models. Equation (\ref{R2b}) is a measure of explained risk in the Cox PH models developed based on entropy loss functions by Heller (2012). Here we call this measure $R_b^2$. The major advantage of the $R_b^2$ measure is that it is compatible to the $R^2$ measure in linear regression models \citep{Heller2012}. The $R^2$ measure in equation (\ref{R2w}) is a simple one to approximate the measure of explained ``randomness" in the Weibull model \citep{Kent1998}, and we call it $R_w^2$.

\section{Simulation Study}
\subsection{Single-Mediator Models}
\subsubsection{Simulation Setting}
Single-mediator models can be described using equations (\ref{eq: X_single}) to (\ref{eq: med_single}) in the Methods section. We assumed the two following true models to generate the data, for subject $i=1,\ldots,n$:
\begin{align*}
& M_i = e + aX_i + \epsilon_i, \\
& \lambda(t|X_i, M_i) = \lambda_0^{(XM)}(t) \exp(rX_i + bM_i).
\end{align*}
\noindent
At the same time, we used the following models together to estimate parameters and compute effect size measures based on the simulated data:
\begin{align*}
& \lambda(t|X_i) = \lambda_0^{(X)}(t) \exp(cX_i), \\
& \lambda(t|M_i) = \lambda_0^{(M)}(t) \exp(dM_i),
\end{align*}
where $X_i$ was an exposure variable, $M_i$ was a mediator, and we assumed that $M_i$ was continuous and linearly regressed on $X_i$. $e$ was an intercept. $c$ was the total effect for the exposure variable, and $\epsilon_i$ was an error term, independent of $X_i$.

We simulated $X_i$ from a standard normal distribution and generated a single mediator $M_i$ from equation (\ref{eq: med_single}). We assumed a common baseline hazard functional form: $\lambda_0(t) = \lambda t^{\lambda - 1} \exp(\eta)$, with $ \lambda = 2$ and $\eta = -5$. The error term $\epsilon_i \overset{i.i.d.}{\sim} N(0,1) $. We generated the survival time $T$ by using equation (\ref{eq: XM_single}).

We further let $ a = 0.5$, $b = -1.5$, $r = 2$, sample size $n=2000$, and replication times $Q=1000$. The indirect mediation effects based on the product and difference measures and $R^2_{med}$ were all calculated by the corresponding functions under the following two settings, based on Cox PH models:

\noindent
{Setting (S1)}: We fixed the sample size $n= 2000$ and let the censor rates vary from 5\% to 95\%, including 5\%, 15\%, 20\%, 25\%, 35\%, 65\%, 85\%, 90\%, and 95\%;

\noindent
{Setting (S2)}: We fixed the censor rate at 25\% and let the sample size very from 200 to 5000, including 200, 500, 1000, 2000, and 5000.

\subsubsection{Simulation Results of Single-Mediator Models}

\begin{figure}[h!]
\begin{center}
\includegraphics[width=12cm,height=18cm]{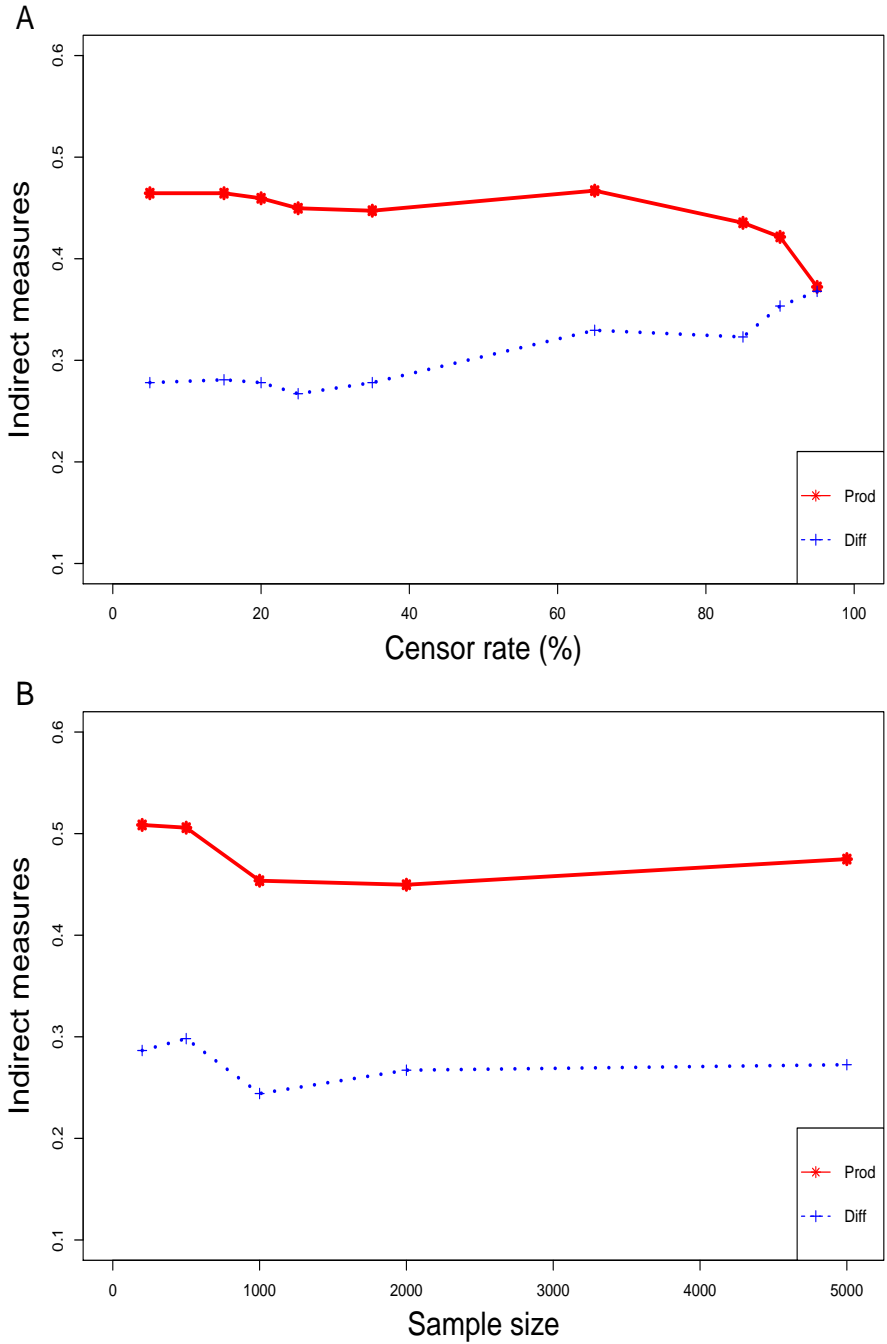}
\end{center}
\caption{Plots of indirect mediation effects by the product (Prod) and difference (Diff) measures under the single-mediator setting. (A) varying the censor rates; (B) Varying the sample sizes.}
\label{f1-1}
\end{figure}
 \cite{Tein2003} reported, without considering censoring and ties, that the product and difference effect size measures of single-mediator analysis were comparable by simulations in log survival time models and log hazard time models, but not in Cox PH models. \cite{VanderWeele2011} provided an analytical approach to prove the equivalence of the two measures for single-mediator analysis in both PH models and accelerated failure time models only when the outcome is rare.

To confirm the above statements, our simulation results showed that the two measures were quite close if the censor rate was above $90\%$; if the censor rate was $95\% $, their values were almost identical, such that they converged as shown in Figure \ref{f1-1}A under the Cox PH modeling framework. If we fixed the censor rate at $85\%$, the two indirect effect measures did converge for varied sample sizes from 50 to 5000, as shown in Figure \ref{f1-1}B. Except for performing mediation analysis under linear regression or non-linear regression (such as Cox PH models) with a rare event, we would get quite different results from these two measures. Therefore, development of more generalized effect size measures are necessary for Cox regression models, and we suggest use of $R^2$ measures.

\begin{figure}[h!]
\begin{center}
\includegraphics[width=15cm,height=14cm]{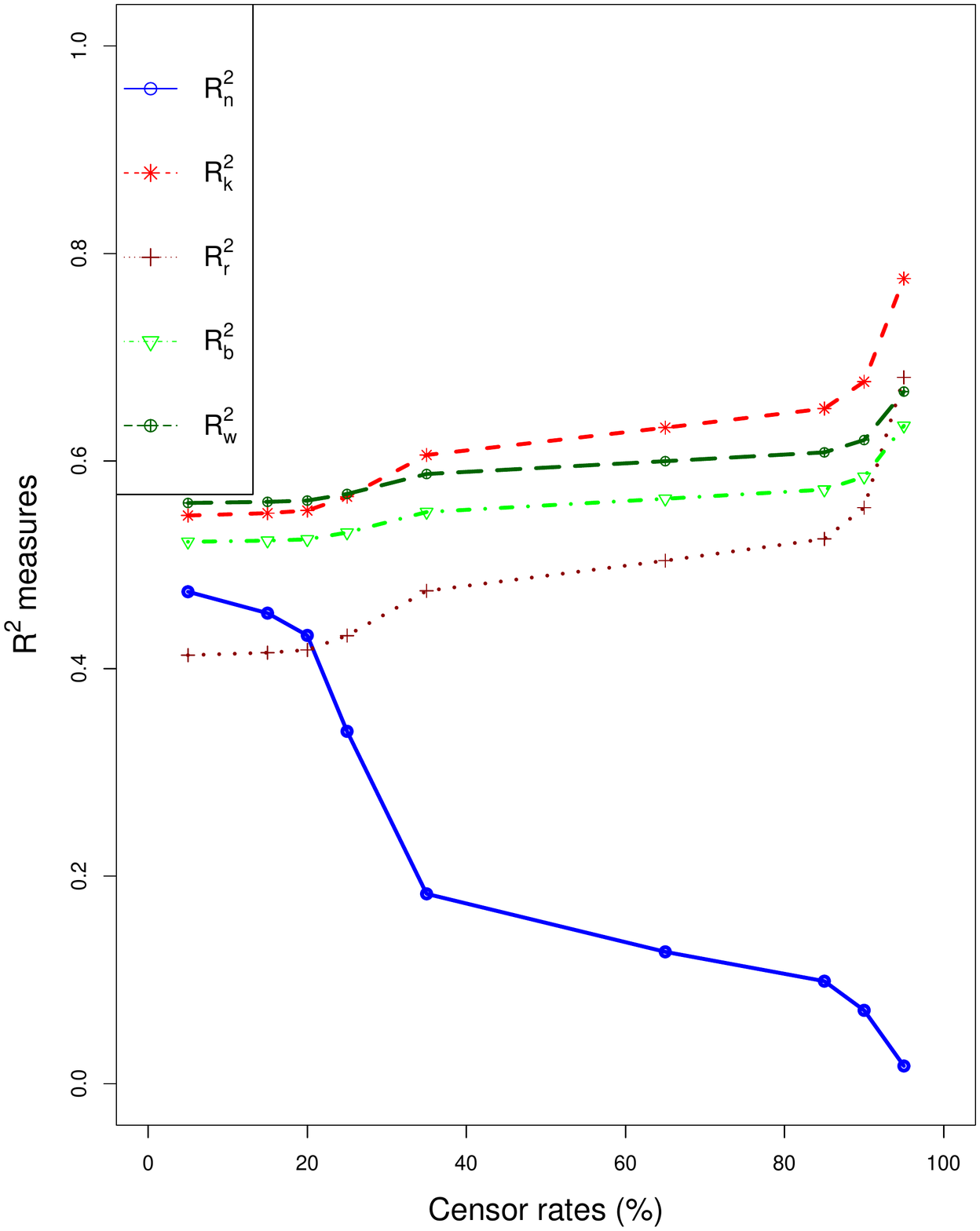}
\end{center}
\caption{Plots of $R^2_{med}$ effect sizes of five different $R^2$ measures by varying censor rates under the single-mediator setting.}
\label{f1-2}
\end{figure}

Under simulation setting (S1), we fixed the sample size and varied censor rates to compare the performance of the five different $R^2$ measures of mediation effect size. As shown in Figure \ref{f1-2}, we found, except for the $R_n^2$ measure, that the rest of the four measures increased slowly with increasing censor rates, going up more dramatically when censor rates reached higher levels (e.g., greater than 90\%). $R_k^2$, $R_b^2$, and $R_w^2$ had relatively closer profiles. $R_k^2$ kept values at a bit of a higher level, and $R_r^2$ kept values lower than $R_b^2$ and $R_w^2$. Therefore, these two measures, $R_b^2$ and $R_w^2$, were relatively independent of censor rate changes. Comparing $R_b^2$ and $R_w^2$ with the product or difference effect size measures under the same simulation setting, these two $R^2$ measures were relatively stable and independent of censoring.

\subsection{Multiple-Mediator Simulations}
\subsubsection{Simulation Setting}
\noindent
Multiple-mediator models under the Cox PH setting follow equations (\ref{eq: X_multiple}) to (\ref{eq: med_multiple}) in the Methods section. As in single-mediator models, we assumed the following two true models to generate the data, for subject $i=1,\ldots,n$ and mediator $j=1,\ldots,p$:
\begin{align*}
& M_{ij} = e_j + a_jX_i + \epsilon_{ij}\\
& \lambda(t|X_i,\boldsymbol{M_{i}}) = \lambda_0^{(XM)}(t) \exp(rX_i + \sum b_j M_{ij}).
\end{align*}
\noindent
We also used the following models together to estimate parameters and compute effect size measures based on the simulated data:
\begin{align*}
& \lambda(t|X_i) = \lambda_0^{(X)}(t) \exp(cX_i), \\
& \lambda(t|\boldsymbol{M_{i}}) = \lambda_0^{(M)}(t) \exp(\sum d_j M_{ij}),
\end{align*}
where $X_i$ was an exposure variable, $M_j$ was a continuous mediator, $e_j$ was an intercept, and $\epsilon_{ij}$ was an error term.  $a_j$, $b_j$, $r$, $d_j$, and $c$ were regression coefficients, among which $c$ was the total effect. $\exp{(r)}$ was the direct effect and $\exp{(a^Tb)}$ was the product indirect effect; these are usually estimated by the partial likelihood approach.


We further let $ a = [a_1, a_2,\cdots,a_p]^T, b = [b_1, b_2,\cdots,b_p]^T $, the sample size $n=2000$, and replication times $Q=1000$, where $p=5$. We assumed a common form of baseline hazard function: $\lambda_0(t) = \lambda t^{\lambda - 1} \exp(\eta)$, a Weibull distribution with $ \lambda = 2$ and $\eta = -5$. We generated the survival time $T$ by using equation (\ref{eq: XM_multiple}). The product and difference measures, along with the five $R^2_{med}$ effect size measures  defined in formula (\ref{R2n} - \ref{R2w}), were all computed via the functions under the following five settings based on Cox PH models.

Setting (M1): We assumed five true mediators existed. The independent variable $X$ followed standard normal distribution $N(0,1)$. The regression coefficients were set as follows: $r=2.5, b_j = 0.5, a_j = 1$, for $j = 1, 2, \cdots, 5$. Therefore, after considering their correlation introduced by the mediated relations among $X$ and all mediators, $(X_i, M_{i1},\ldots, M_{i5})^{T}$ were generated altogether by a multivariate normal distribution with a mean vector of $0$ and a variance-covariance matrix with a compound symmetry structure: its diagonal was a vector $(1, 1.25, 1.25, 1.25, 1.25, 1.25)^{T}$, and all off-diagonal values were 0.25.  We varied the censor rates at 5\%, 25\%, 65\%, 85\%, and 95\%.\\
\noindent
Setting (M2): We assumed five true mediators existed. The independent variable $X$ and all mediators followed a multivariate normal distribution as in setting (M1), but its variance-covariance matrix changed with varying values of $a$. We varied $a$ by letting $a_j = 0.05,0.25,0.5,1,3$, or $5$, and the rest of the regression coefficients were set as follows: $r=2.5$ and $b_j = 0.5$, for $j=1,\ldots,5$. The censor rate was set at 85\%.\\
\noindent
Setting (M3): The independent variable $X$ and all mediators followed multivariate normal distribution as in setting (M1). We varied $b$ by letting $b_j = 0.01,0.25,0.5,1,2$ ,or $3$, and the rest of the regression coefficients were set as follows: $r=2.5$  and $a_j = 1$, for $j=1,\ldots,5$. The censor rate was set at 85\%.\\
\noindent
Setting (M4): The independent variable $X$ and all mediators followed multivariate normal distribution as in setting (M1). We varied the direct effect $r$ by letting $r= 0.05,0.5,1,2,5$, or $10$, and the rest of the regression coefficients were set as follows: $ b_j = 0.5; a_j = 1$ for $j=1,\ldots,5$. The censor rate was set at 85\%.\\
\noindent
Setting (M5): The independent variable $X$ and all mediators followed multivariate normal distribution as the setting (M1), and the censor rate was set at 85\%. We varied the number of mediators from 1 to 5 to assess the mediation effects for nested models.

\subsubsection{Simulation Results of Multiple-Mediator Models}

\begin{figure}[h!]
\begin{center}
\includegraphics[width=14cm,height=18cm]{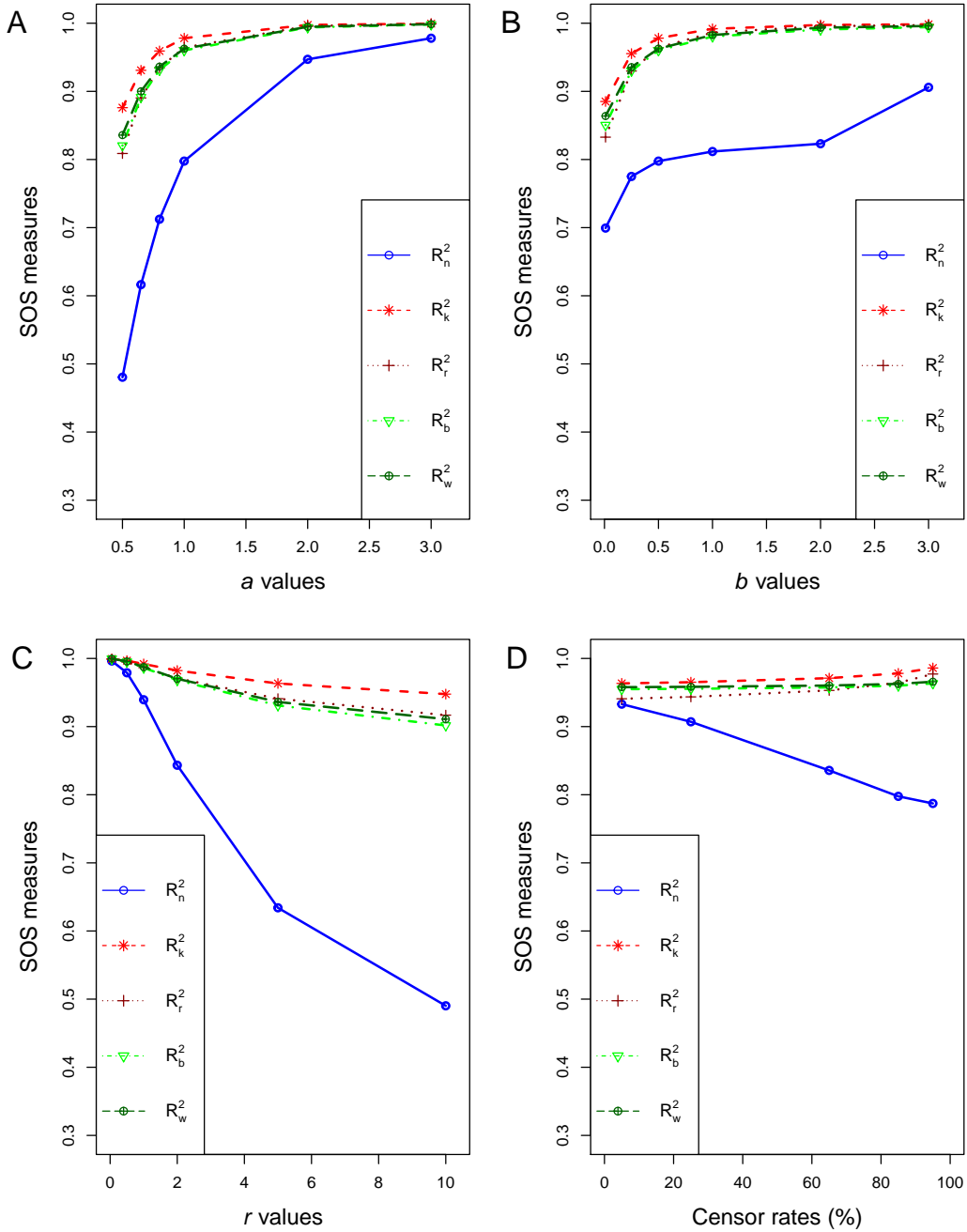}
\end{center}
\caption{Plots of SOS mediation effects for five different $R^2$ measures by varying strength of association and censor rates. (A) Varying $a$ values; (B) Varying $b$ values;(C) Varying $r$ values; (D) Varying censor rates.}
\label{f2}
\end{figure}
\noindent
We first studied the properties of $R^2$ measures. Under the above five simulation settings of multiple-mediator models, we compared the performance of the five different $R^2$ measures of mediation effect size, and investigated whether any of them could follow the desirable characteristics of good $R^2$ measures listed in Section \ref{R2prop}.

From simulation results shown in Figure \ref{f2}D, we found the $R_n^2$ measure decreased with increasing censor rates, and the other four measures were relatively independent of censor rates (property(b)). By varying $a$ and $b$, shown in Figure \ref{f2}A and B, we could see the mediation effects were increasing with the growth of coefficient $a$ and $b$ values. By varying $r$, shown in Figure \ref{f2}C, we could see the opposite trend: the mediation effects were decreasing with an increase in coefficient $r$ values. In all these settings, the $R_n^2$ measure changed more dramatically relative to the other $R^2$ measures, and its values were much lower than the other four. The $R_k^2$ measure was always relatively higher than the other three, i.e., $R_r^2$, $R_b^2$, and $R_w^2$, which almost overlapped with each other. Therefore, we can interpret that these three measures of mediation effects were increasing with an increase in the effect size of $a$ or $b$, except at the highest level (e.g., such as $a$ or $b$ values reaching from 2 to 3, which were too high and unrealistic values). Similarly, when we varied $r$ from 0.05 to 10, within the reasonable range from 0.05 to 6, all three $R^2$ SOS measures were decreasing more reasonably than the other two measures.  So all of these results may reflect that the three measures $R_r^2$, $R_b^2$ and $R_w^2$ satisfied property (c).

\begin{figure}[h!]
\begin{center}
\includegraphics[width=12cm,height=14cm]{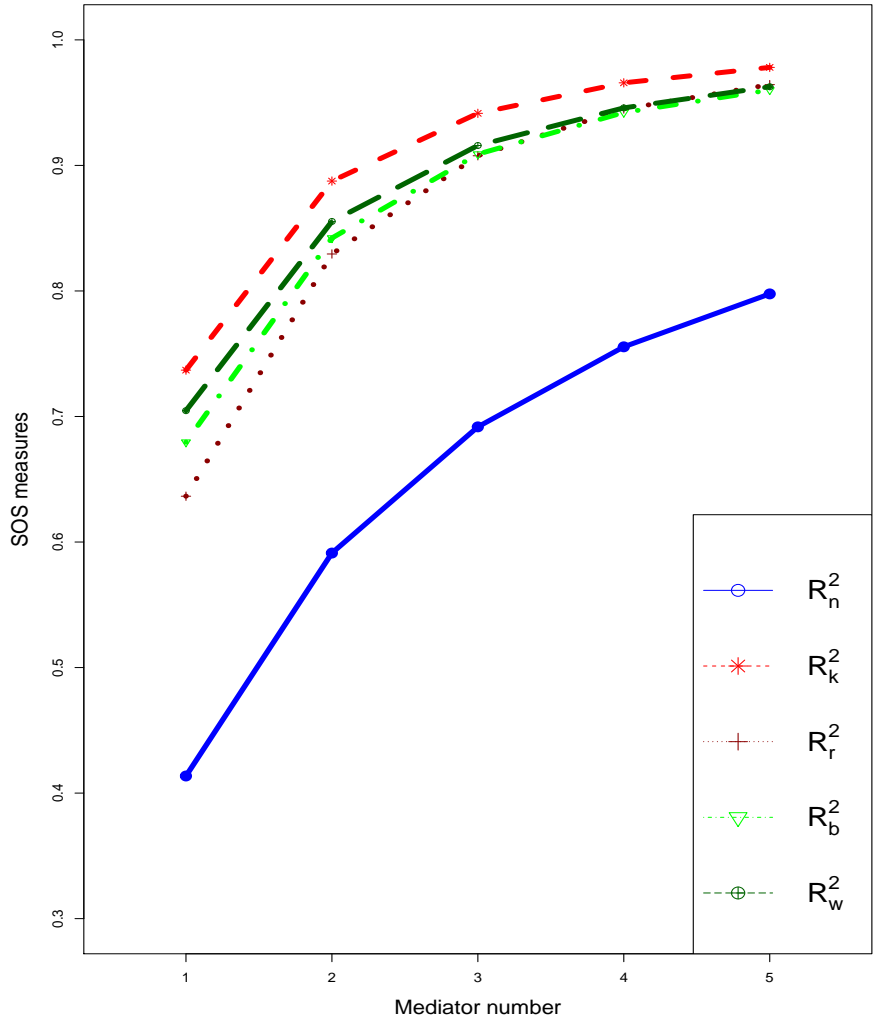}
\end{center}
\caption{Plots of five different $R^2$ SOS measures by varying the number of mediators from 1 to 5.}
\label{f3}
\end{figure}

We also investigated the nested property by varying the number of mediators. As shown in Figure \ref{f3}, all five $R^2$ SOS measures increased with the increasing number of mediators from 1 to 5. The only difference was that $R_n^2$ increased a bit faster, and at the same time, its values were much lower than the other four measures. $R_n^2$ kept relatively higher values, and $R_r^2$ kept relatively lower ones compared with $R_b^2$ and $R_w^2$. So these results may reflect that all $R^2$ SOS measures satisfied property (d).

Both $R_b^2$ and $R_w^2$ were closely related to explained variation, which is ``equivalent" to linear regression analysis (property(a)). Their formula also suggested some flexibility to change with model structure alterations, such as adding covariates or considering  interaction terms \citep{Royston2006, Heller2012}. Their confidence interval can be approximately calculated using the delta method or bootstrap approach (property(e)). Therefore, in conclusion, we recommend that the $R^{2}_{med}$ and $SOS$ based on $R_b^2$ and $R_w^2$ are better effect size measures for multiple-mediator models under the Cox PH framework.

\noindent

\begin{figure}[h!]
\begin{center}
\includegraphics[width=14cm,height=18cm]{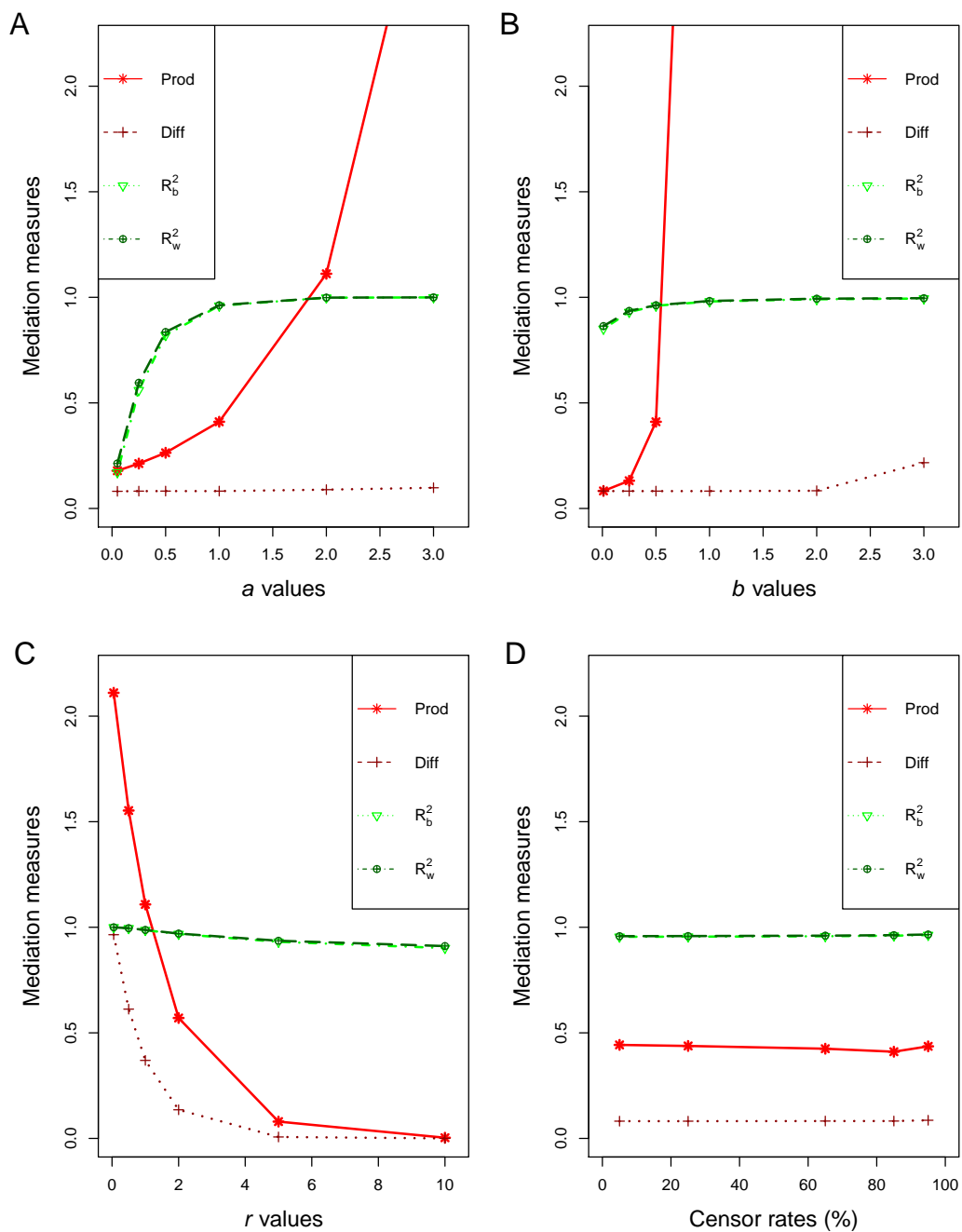}
\end{center}
\caption{Plots of mediation effects comparing two $R^2$-based SOS measures with the product (Prod) and difference (Diff) indirect effect measures. (A) Varying $a$ values; (B) Varying $b$ values;(C) Varying $r$ values; (D) Varying censor rates. Note: $R^2_b$ and $R^2_w$ curves almost overlap in plots from (A) to (D).}
\label{f4}
\end{figure}

We compared the $R^2$ SOS measures, $R_b^2$ and $R_w^2$, with the product and difference indirect effect measures, which are commonly used under the single-mediator setting. We calculated the proportion mediated indirect effects for these two measures under varying censor rates and strength of association in our multiple-mediator models. From the simulation results shown in Figure \ref{f4}, we can see that for all the settings, the difference measure was hardly sensitive to varying values of $a$ and $b$ compared with the $R^2$ measures. On the other hand, the product measure under varying strengths of association could change dramatically and be totally off. With varying censor rates, the patterns were similar for all four measures. Therefore, the difference measure compared with the product measure was more similar to $R^2$ measures except in a much smaller value range. Thus, such comparisons support our suggestion to use the $R^2$ measures for mediation effect.

\noindent
\section{Real Data Example: Application to the Framingham Heart Study}
\subsection{Dataset Description}
The Framingham Heart Study, which began in 1948, is a family-based, ongoing longitudinal prospective cohort study \citep{Splansky2007}. It primarily investigates cardiovascular diseases among the residents of the town of Framingham, Massachusetts. This study has included participants from three generations: the original cohort (first generation), the offspring cohort (second generation), and the third generation cohort. The offspring cohort is the largest one. For our present study, we used the offspring cohort dataset, and it included both longitudinal phenotype measurements and time to disease events for 1523 unrelated participants. We downloaded the data from the US National Center for Biotechnology Information dbGaP \citep{Cao2016}.

We were interested in two survival outcomes in the FHS data: CHD occurrence and all-cause mortality. It has been reported that the following major risk factors affect CHD events: lipid abnormalities, cigarette smoking, drinking, blood pressure, older age, and obesity (body mass index (BMI) 30 or higher) \citep{Wilson1998}. Such risk factors are also associated with general or coronary cause-specific mortality \citep{ Sharif2019,Keeley2010}. In general, cholesterol is a fat-like substance in the blood, which is necessary for building-up of arteries walls. However, high blood cholesterol is a major risk factors for heart disease: the higher the blood LDL-C level, the greater the risk for developing heart disease or having a heart attack. On the other hand, a higher HDL-C level can help to clear an overabundance of cholesterol in the blood, benefiting health. These risk factors are in generally highly correlated. Previous research suggest that increased smoking or smoking paired with drinking and other factors may be responsible for increases in TC, LDL-C, and TG, as well as decreases in HDL-C \citep{Gossett2009, Mukamal2006, Vodnala2012}. Therefore, several or all risk factors may work together to confer the risk of heart diseases \citep{Buttar2005, Jousilahti1996}. To this end, we would like to untangle this research ``puzzle" in order to understand why and how these factors work together to contribute to the occurrence of clinical outcomes by mediation analysis.

Specifically, we used three variables together as exposures, including current smoking, current drinking, and gender. We used the baseline age as a left truncation variable. The mediators were HDL-C, LDL-C, TC, TG, glucose, and BMI. We evaluated their mediation effects via SOS of $R^2$ measures in the single and multiple-mediator models with CHD risk and the risk of all-cause mortality. Among the 1523 subjects, there were 209 incident CHD events and 206 all-cause deaths with a censoring rate of 86.3\% and 86.4\%, respectively. The follow-up duration ranged from 8 years to 44 years with a median of 41 years. The results are summarized in Tables \ref{T1} and \ref{T2}.

\subsection{Single-Mediator Analysis of the FHS Data}
In order to show the conformity of results from our mediation analyses using single-mediator and multiple-mediator models, we first evaluated all six mediators one-by-one in single-mediator models for both CHD risk and all-cause mortality risk. From Table \ref{T1}, we found that HDL-C could explain the highest variation of both CHD risk and all-cause mortality risk in all five $R^2$ measures. 
However, the $R^2_{med}$ measure might be negative in single-mediator models, which could be explained for two reasons. First, the mediation effect size was really small, or close to zero. For instance, we simulated an unrelated variable (called ``Random"), and its $R^2$ values were small around 0, which sometimes lead to $R^2_{med}$ negativity. Second, if the directions of the indirect and direct effects were different (for example, TC in CHD risk or LDL-C in all-cause mortality), this might result in $R^2_{med}$ negativity, consistent with other reports \citep{Yang2019}. We will expand this in the Discussion section.

\begin{table}
\caption{\label{T1} SOS evaluation of $R^2_{med}$ measures in single-mediator models}
\centering
\begin{threeparttable}
\begin{tabular}{p{2cm} p{2cm} p{2cm} p{2cm} p{2cm} p{2cm}}
\hline
\multicolumn{6}{l}{Time to CHD} \\
\hline
Mediators & SOS-$R^2_n$ & SOS-$R^2_k$ & SOS-$R^2_r$ & SOS-$R^2_b$ & SOS-$R^2_w$\\
\hline
BMI & 0.14 &  0.16 & 0.14 & 0.16 & 0.14\\
Fgluc & 0.27 &  0.35 & 0.29 & 0.32 & 0.23\\
HDL-C & 0.68 &  0.73 & 0.69 & 0.85 & 0.83\\
LDL-C & 0.070 &  0.12 & 0.079 & 0.088 & 0.078\\
TC & -0.13 &  -0.10 & -0.13 & -0.16 & -0.12\\
TG & 0.35 &  0.50 & 0.39 & 0.70 & 0.32\\
Random & -0.0057 &  0.0017 & -0.0046 & 0.0053 & 0.0067\\
\hline
\multicolumn{6}{l}{Time to all-cause mortality} \\
\hline
BMI & 0.011 &  0.019 & 0.012 & 0.010 & 0.0073\\
Fgluc & 0.025 &  0.072 & 0.045 & 0.026 & 0.047\\
HDL-C & 0.14 &  0.16 & 0.15 & 0.14 & 0.18\\
LDL-C & -0.0056 &  -0.012 & -0.049 & -0.046 & -0.039\\
TC & 0.048 &  0.053 & 0.048 & 0.056 & 0.067\\
TG & 0.048 &  0.055 & 0.048 & 0.039 & 0.039\\
Random & 0.011 &  0.021 & 0.012 & 0.023 & 0.029\\
\hline
\end{tabular}
\begin{tablenotes}
\item BMI: body mass index; Fgluc: fasting glucose; HDL-C: high density lipoprotein cholesterol; LDL-C: low density lipoprotein cholesterol; TC: total cholesterol; TG: triglycerides; Random: a simulated random variable. SOS: the shared over simple effect.
\end{tablenotes}
\end{threeparttable}
\end{table}

\subsection{Multiple-Mediator Analysis of the FHS Data}

\begin{table}
\caption{\label{T2} SOS evaluation of $R^2_{med}$ measures in multiple-mediator models}
\centering
\begin{threeparttable}
\begin{tabular}{p{2cm} p{2cm} p{2cm} p{2cm} p{2cm} p{2cm}}
\hline
\multicolumn{6}{l}{T = Time to CHD }  \\
\hline
$R^2$ Type & $R^2(T,X)$ & $R^2(T,M)$ & $R^2(T,XM)$ & $R^2_{med}$ & SOS\\
\hline
$R_n^2$ & 0.024 &  0.080 & 0.085 & 0.019 & 0.81\\
$R_k^2$ & 0.16 &  0.45 & 0.47 & 0.14 & 0.88\\
$R_r^2$ & 0.10 &  0.34 & 0.35 & 0.088 & 0.84\\
$R_b^2$ & 0.12 &  0.50 & 0.49 & 0.13 & 1.055\\
$R_w^2$ & 0.15 &  0.33 & 0.35 & 0.13 & 0.87\\
\hline
\multicolumn{6}{l}{T= Time to all-cause mortality } \\
\hline
$R_n^2$ & 0.031 &  0.013 & 0.042 & 0.0014 & 0.047\\
$R_k^2$ & 0.21 &  0.091 & 0.27 & 0.027 & 0.13\\
$R_r^2$ & 0.14 &  0.057 & 0.18 & 0.0098 & 0.072\\
$R_b^2$ & 0.21 &  0.061 & 0.25 & 0.018 & 0.086\\
$R_w^2$ & 0.20 &  0.080 & 0.25 & 0.027 & 0.13\\
\hline
\end{tabular}
\begin{tablenotes}
\item X(exposures): gender, smoking, and drinking status; M (mediators): body mass index, fast glucose, high density lipoprotein cholesterol, low density lipoprotein cholesterol, total cholesterol, and triglycerides; SOS: the shared over simple effect.
\end{tablenotes}
\end{threeparttable}
\end{table}

\noindent
From the results of multiple-mediator models in Table \ref{T2}, we found that, except for $R^2_{n}$, the exposure variables X that included gender, smoking, and drinking status could explain 10-16\% of the variation in CHD risk and 14-21\% of the variation in all-cause mortality risk. All six mediators together explain 33-50\% of the variation in CHD risk and 5.7-9.1\% of the variation in all-cause mortality risk. All six mediators plus exposures could explain 35-49\% of the variation in CHD risk and 18-27\% of the variation in all-cause mortality risk. After calculation of the mediation effect at the SOS scale for all five $R^2$ measures, six mediators altogether could explain more than 80\% of the variation in CHD risk from exposure variables. $R_b^2$ provided the highest value (1.05), which makes sense for two reasons. First, here we used almost all known CHD risk factors as our mediators to explain CHD risk variation in our exposure variables, including gender, drinking, and smoking. Second, $R^2_{med}$ is not bounded in the interval (0, 1), thus SOS may be larger than 1 if correlation among mediators or exposure variables is high. On the other hand, the six mediators were able to explain about 5\% - 14\% of the all-cause mortality risk from the exposure variables in all five $R^2$ measures under the same settings. This suggests that these six well-established lifestyle risk factors for cardiovascular diseases mediated much less on the exposure-all-cause-mortality association than the exposure-CHD association.

\section{Discussion and Conclusion}

There are multiple challenges when estimating mediation effects. For example, we have to choose the right effect-size measure, especially under a Cox regression setting. Multiple or high-dimensional mediators are often encountered in practice. Thus, our goal in this article was to study the statistical characteristics and evaluate the performance of $R^2_{med}$ effect size measures for multiple-mediator models under the Cox PH regression framework. We would like to give a reasonable solution to the posed challenges except implying causal inference. Since causal mediation models require very strong assumptions, which may be often violated and difficult to verify in real data applications. Therefore, we do not pursue cause-and-effect interpretation in this article.

We extended mediation analysis from single-mediator and linear regression based models to multiple-mediators and Cox PH regression models. At the same time, we chose the $R^2$ measure instead of the commonly used product measure of indirect mediation effect. The chosen $R^2$ measure also posed another layer of challenges to mediation analyses under Cox regression frameworks due to the multiple definitions of $R^2$ for survival outcomes \citep{HonerkampSmith2016, ChoodariOskooei2012a, ChoodariOskooei2012b}. According to the five properties of a good $R^2$ measure by Royston's suggestion under Cox regression models, we chose five $R^2$ measures: $R^2_n, R^2_k, R^2_r, R^2_b$, and $R^2_w$ to evaluate their performance in our simulation studies and real data analyses. $R^2$ measures, in general, are stable and easier to extend to multidimensional or even higher-dimensional models compared to commonly used mediation effect measures \citep{Yang2019}. By simulation studies, the $R^2_{med}$ mediation effect measures based on $R_b^2$ and $R_w^2$ are suggested as better for multiple-mediator analysis of Cox PH models due to their satisfying all five properties of \cite{Royston2006}. By using the recommended measures in the FHS data, we found that all six major mediators together including the four lipids, BMI, and glucose, were able to explain 80\% - 90\% of the variation in CHD risk from the exposure variables, and explain the 4\% - 14\% of the variation in all-cause mortality risk from the same set of exposure variables. Due to lack of multiple-mediator models under similar settings, it is hard to compare our results with others. However, these results were consistent with those of our single-mediator models.

Although we obtained favorable results in the $R^2_{med}$ measures, the present study has three possible limitations.  First, we assumed that the mediation models were correctly specified; the performance of the $R^2_{med}$ measure in missepecified models warrants future investigation. Second, more complicated model structures, such as interactions among exposure variables and mediator variables were not included in our studies. So the $R^2_{med}$ measure was not intended to extend for such complicated models. Third, from the results of single-mediator models in real data analyses, as well as its definition (formula (\ref{eq: R2medPH})), we could see that $R^2_{med}$ was not bounded. It can be negative or above 1. Such a result could limit its interpretation.

The major reasons of negativity are as follows: if the indirect mediation effect of a mediator is very small (close to zero),  a $R^2_{med}$ measure is negative due to randomness in data. Another reason is the ``masking" effect among exposure variables and mediators with respect to their outcomes \citep{deHeus2012, Yang2019}. It is very likely that mediators or exposures have different directions of indirect or direct effects,
resulting in the negativity of $R^2_{med}$. However, as the number of mediators increases, the number of possible combinations of the directions of the mediation effects increase \citep{MacKinnon2008, Yang2019}. In fact, \cite{Yang2019} showed that, under the high-dimensional mediator setting, the negativity of $R^2_{med}$ is unlikely to occur, which is consistent with the results of our simulation study and real data analysis (Table \ref{T2}).

In summary, our models extend from the linear, single-mediator analysis for the following three aspects: 1) from univariate to multiple mediators ($P \le 10$); 2) our model formulation and estimation apply under the Cox PH regression framework; 3) we suggested using the two generalized $R^2$ measures: $R^2_b$ and $R^2_w$ in Cox regression-based mediation analysis.

\section*{Acknowledgments}
This research was supported by the National Institutes of Health (NIH) grant R01HL1167202; P.W. was partially supported by NIH grants R01CA16912 and P50CA21767; X.H. was partially supported by the National Science Foundation through grant DMS 1612965, U54CA096300, U01CA152958 and 5P50CA100632, and the Dr. Mien-Chie Hung and
Mrs. Kinglan Hung Endowed Professorship.  The Framingham Heart Study is conducted and supported by the National Heart, Lung, and Blood Institute (NHLBI) in collaboration with Boston University (Contract No. N01-HC-25195). This manuscript was not prepared in collaboration with investigators of the Framingham Heart Study and does not necessarily reflect the opinions or views of the Framingham Heart Study, Boston University, or NHLBI. The authors acknowledge the Texas Advanced Computing Center at The University of Texas at Austin for providing HPC resources and thank Ms. Jessica Swann for editorial assistance. The authors declare no conflict of interest.

\bibliographystyle{rss}
\bibliography{SurvivalMediationR2}
\end{document}